\documentclass[fleqn,10pt]{wlscirep}
\usepackage{graphicx}
\usepackage{bm}

\newcommand{\beq}{\begin{equation}}
\newcommand{\eeq}{\end{equation}}
\newcommand{\beqa}{\begin{eqnarray}}
\newcommand{\eeqa}{\end{eqnarray}}

\newcommand{\kBT}{\mbox{$k_{\rm B}T$}}
\newcommand{\TR}{\mbox{$T_{\rm R}$}}

\renewcommand{\vec}[1]{\mbox{\boldmath$#1$}}

\title{Crossover from BKT-Rough to KPZ-Rough Surfaces for Interface-Limited Crystal Growth/Recession}

\author[1,*]{Noriko Akutsu}
\affil[1]{Faculty of Engineering, Osaka Electro-Communication University, Hatsu-cho, Neyagawa, Osaka 572-8530, Japan}

\affil[*]{nori3@phys.osakac.ac.jp}


\begin{abstract}

The crossover from a Berezinskii--Kosterlitz--Thouless (BKT) rough surface to a Kardar--Parisi--Zhang (KPZ) rough surface on a vicinal surface is studied using the Monte Carlo method in the non-equilibrium steady state in order to address discrepancies between theoretical results and experiments.
The model used is a restricted solid-on-solid (RSOS) model with a discrete Hamiltonian without surface or volume diffusion (interface limited growth/recession).
The temperature, driving force for growth, system size, and surface slope dependences of the surface width are calculated for vicinal surfaces tilted between the (001) and (111) surfaces.
The surface velocity, kinetic coefficient of the surface, and mean height of the locally merged steps are also calculated. 
In contrast to the accepted theory for (2+1) surfaces, we found that the crossover point from a BKT (logarithmic) rough surface to a KPZ (algebraic) rough surface is different from the kinetic roughening point for the (001) surface. The driving force for crystal growth was found to be a relevant parameter for determining whether the system is in the BKT class or the KPZ class. It was also determined that ad-atoms, ad-holes, islands, and negative-islands block surface fluctuations, which contributes to making a BKT-rough surface. 
\end{abstract}
\begin{document}

\flushbottom
\maketitle
%
%



\section*{Introduction}

Surface roughness \cite{bcf,weeks73} is important both practically, in the theory of crystal growth, and fundamentally, in the basic theory of interface properties. 
At equilibrium, the Berezinskii--Kosterlitz--Thouless (BKT) \cite{berezinskii71,  kt} roughening phase transition \cite{chui76, knops, beijeren77, weeks80} occurs at the roughening temperature $\TR$ on a two-dimensional (2D) low-Miller-index surface (interface), such as the (001) surface, in 3D.
For temperatures $T \geq \TR$, the square of the surface width diverges logarithmically as the linear system size $L$ increases to infinity (BKT-rough surface). 

For scaling, the concept of the self-affine surface (interface) has been successful and has been widely used \cite{barabasi, krug91, takeuchi13, takeuchi18}.
The surface width shows a Family--Viscek scaling relation and the surface width $W(L,t)$ can be expressed by the following relation:
\beq
W(L,t)\sim L^\alpha f(L^{-z}t), \ z=\alpha/\beta,  \label{eq_scaling}
\eeq
where  $t$ is time and the $\alpha$, $\beta$, and $z$ exponents are referred to as the roughness, growth, and dynamic exponents, respectively. 

The surface growth equation with a non-linear term under a symmetry principle consideration was first proposed by Kardar, Parisi, and Zhang (Kardar--Parisi--Zhang, KPZ) \cite{kpz}.
For a two-dimensional (2D) surface in 3D, the exponents are obtained numerically as $\alpha=0.3869$, $\beta= 0.2398$, and $z=1.6131$ \cite{pagnani15, takeuchi18} (KPZ-rough surface).
The values of the exponents have been observed for directed polymers, as well as other systems in the KPZ universality class.


\begin{figure}
\centering
\includegraphics[width=5.0 cm,clip]{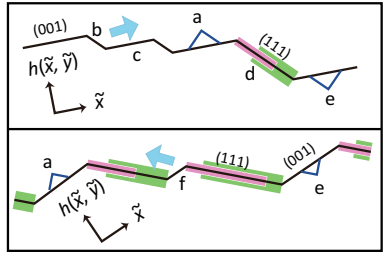}%
\caption{
Schematic figures of the side views of vicinal surfaces in the RSOS model.
Dark lines: profile of the surfaces.
Pink lines: part of the local (111) surface where ad-atoms cannot be added.
Green thick lines: part of the local (111) surface where ad-holes cannot form. 
Upper box: vicinal surface tilted from the (001) surface.
Lower box: vicinal surface tilted from the (111) surface.
Thick light-blue arrows: step-growth direction for crystal growth.  
$\tilde{x}$ and $\tilde{y}$ indicate the $\langle 110 \rangle$ and $\langle \bar{1}10 \rangle$ directions.
a: ad-atom.  b: mono-atomic step.  c: (001) terrace.
d: step with a height of three mono-atomic steps ; the side surface is (111) surface.
e: ad-hole (negative ad-atom).
f: negative step.  
}  
\label{schem}
\end{figure}

However, for crystal growth, the experimentally observed exponents are typically different from the KPZ exponents \cite{barabasi, krim95, villain91, pimpinelli}.
The question of the reason for the difference between KPZ growth and the experimentally observed crystal growth has attracted considerable attention \cite{takeuchi18, xia20, einstein99,  pimpinelli02, pimpinelli}.
For crystal growth with surface diffusion, step-flow growth \cite{bcf,villain91, pimpinelli} on a vicinal surface is expected.
A vicinal surface at temperatures less than $\TR^{(001)}$ can be described by terrace surfaces and a train of steps, where a step consists of a zig-zag structure on the edge (the terrace, step, kink (TSK) picture, Fig. \ref{schem}) \cite{bcf}.
At equilibrium, the square of the surface width of a vicinal surface diverges logarithmically as the system size diverges  \cite{huse85, dennijs85, akutsu87, yamamoto94, akutsu94} for $T<\TR$, similar to a BKT-rough surface.
This logarithmic divergence results from long wavelength slope fluctuations \cite{dennijs85, akutsu87} caused by step-wandering \cite{huse85}.

In the non-equilibrium state, one reason why the crystal surface is different from the KPZ class \cite{villain91,pimpinelli} is surface diffusion, such as in the case of molecular beam epitaxy (MBE). 
The surface width shows algebraic divergence but with different values of the exponents from those of KPZ. 
Depending on the step--step interactions, several groups of exponents are obtained theoretically \cite{pimpinelli02, tonchev16}.

Recently, for reaction-limited crystal growth, different exponents were experimentally obtained \cite{gupta16} from the KPZ values. 
In addition, in the solution growth of SiC and GaN, self-assembled faceted macrosteps roughen the vicinal surface and degrade the quality of the crystal \cite{mitani}.
In our previous work, we used the restricted solid-on-solid (RSOS) model with a point-contact-type step--step attraction (p-RSOS model) to demonstrate that a faceted macrostep exists stably at equilibrium \cite{akutsu09, akutsu11JPCM,akutsu15book,akutsu16,akutsu16-3}.
Here, ``restricted'' means that the surface height difference between nearest neighbor sites is restricted to $\{0, \pm 1\}$.
For interface-limited crystal growth/recession, a faceted macrostep disassembles as the absolute value of the driving force for crystal growth increases \cite{akutsu17,akutsu18,akutsu19,akutsu19-2}, which is a different behaviour than that of the results obtained for diffusion limited crystal growth \cite{chernov61}.

Hence, in this article, the crossover from a BKT-rough surface to a KPZ-rough surface for {\it interface-limited growth} is studied using the Monte Carlo method based on the RSOS model \cite{rsos,dennijs,yakutsu89} with a discrete Hamiltonian equivalent to the 19-vertex model.
The surface is tilted between the (001) surface and the (111) surface. 
The surface width, surface velocity, and mean height of the locally merged steps are calculated depending on a set of external parameters, temperature $T$, driving force for crystal growth $\Delta \mu$, linear size of the system $L$, and surface slope $p$ for the interface limited growth (recession) in the non-equilibrium steady state.
The calculated results for the  slope dependence of the surface width, surface velocity, and mean height of the locally merged steps are of greatest interest.
From these results, we demonstrate which parameter determines whether the surface is KPZ-rough or BKT-rough.
This work builds a bridge between mathematical models and surface models for crystal growth.

It should be noted that the RSOS model applied in the present study is slightly different from the RSOS model studied by Kim and Kosterlitz \cite{kim89}, which corresponds to the absolute SOS (ASOS) or simply the SOS model \cite{muller-krumbhaar78} for crystal growth, where the height difference between nearest neighbour sites can take a natural number up to the linear system size normal to the surface.
However, their numerical simulations based on the KPZ equation were performed for a height difference up to 1. 
The crossover from a BKT-rough surface to a KPZ-rough surface for a 2D vicinal surface in 3D was first discussed by Wolf \cite{wolf91} using renormalization calculations with the anisotropic KPZ (AKPZ) equation.
However, the present results are different from  the AKPZ results on some points.

To obtain clear results for interface-limited crystal growth/recession, the surface diffusion \cite{bcf,villain91,pimpinelli}, volume diffusion \cite{chernov61}, 
second-nearest-neighbour interaction between atoms \cite{krzy14} in crystals, Ehrlich--Schwoebel  effect \cite{ehrlich, Schwoebel}, elastic interactions \cite{alerhand}, surface reconstruction \cite{williams93,yamamoto94, akutsu94}, adsorption effects \cite{akutsu01,akutsu01-3, akutsu03, akutsu09-2}, and point-contact-type step--step attraction \cite{akutsu09, akutsu11JPCM, akutsu15book, akutsu16, akutsu16-3} are not taken into consideration.

\section*{Model and Calculations}


%
The Hamiltonian for a vicinal surface is given by the following equation:
\beq
{\mathcal H}= \sum_{\{m,n\}} \left\{\epsilon[|h(m+1,n)-h(m,n)| \right. 
+
 |h(m,n+1)-h(m,n)|] 
-
\Delta \mu  \ h(m,n)\} 
+ {\mathcal N} E_{{\rm surf}}, 
\label{hamil}
\eeq 
where $h(m,n)$ is the height of the surface at a site $(n,m) $, $\epsilon$ is the microscopic ledge energy, ${\mathcal N}$ is the total number of unit cells on the (001) surface, and $E_{{\rm surf}}$ is the surface energy per unit cell.
The RSOS condition is required implicitly.
Here, $\Delta \mu$ is introduced such that $\Delta \mu= \mu_{\rm ambient} - \mu_{\rm crys}$, where $\mu_{\rm ambient}$ and $\mu_{\rm crys}$ are the bulk chemical potential of the ambient and crystal phases, respectively.
At equilibrium, $\Delta \mu=0$; for $\Delta \mu>0$, the crystal grows; whereas for $\Delta \mu<0$, the crystal recedes.
The (grand) partition function for the surface at equilibrium is obtained by $Z(T, L, \Delta \mu, N_{\rm step})|_{\Delta \mu=0} = \sum_{h(m,n)} \exp[ - {\cal H}/\kBT]$ with a fixed $N_{\rm step}$.

For first-principles quantum mechanical calculations, $E_{{\rm surf}}$ or $\epsilon$   corresponds to the surface free energy, which includes the entropy originating from lattice vibrations and distortions \cite{kempisty19}.
Hence, $E_{{\rm surf}}$ or $\epsilon$ decreases slightly as the temperature increases.  
However, $E_{{\rm surf}}$ and $\epsilon$ are assumed to be constant throughout this work because we concentrate on the crossover phenomena of the surface roughness.


The vicinal surfaces of the tilted (001) and (111) surfaces are considered by using the Monte Carlo method with the Metropolis algorithm.
Atoms are captured from the ambient phase to the crystal surface, and escape from the crystal surface to the ambient phase.
The number of atoms in a crystal is not conserved.
The external parameters are temperature $T$, $\Delta \mu$, number of steps $N_{\rm step}$, and the linear size of the system $L$.
The (mean) surface slope $p$ is defined by $p = \tan \theta = N_{\rm step}a/L$.
For details of the Monte Carlo calculations, refer to Ref. \cite{akutsu20} and the Supplementary Information.


The square of the surface width $W(L,t)$ is  defined by the variance of the height $h(\vec{x},t)$ of the vicinal surface:
\beqa
gW(L,t)^2&=&\langle [h(\vec{x},t) - \langle h(\vec{x},t) \rangle]^2 \rangle, 
\nonumber \\ 
g &=& (1 + p_x^2 + p_y^2) = 1/\cos ^2 \theta, \quad 
p_x =  p_y= N_{\rm step}a\sqrt{2}/L,
\label{eq_def_w2}
\eeqa
where $\vec{x}$ is a site on the surface, $g$ is the determinant of the first fundamental quantity of a curved surface \cite{kreyszig, akutsu87}, and $\theta$ is the tilt angle inclined towards the $\langle 111 \rangle$ direction from the $\langle 001 \rangle$ direction.

\section*{Results}


\begin{figure}
\centering
\includegraphics[width=11 cm,clip]{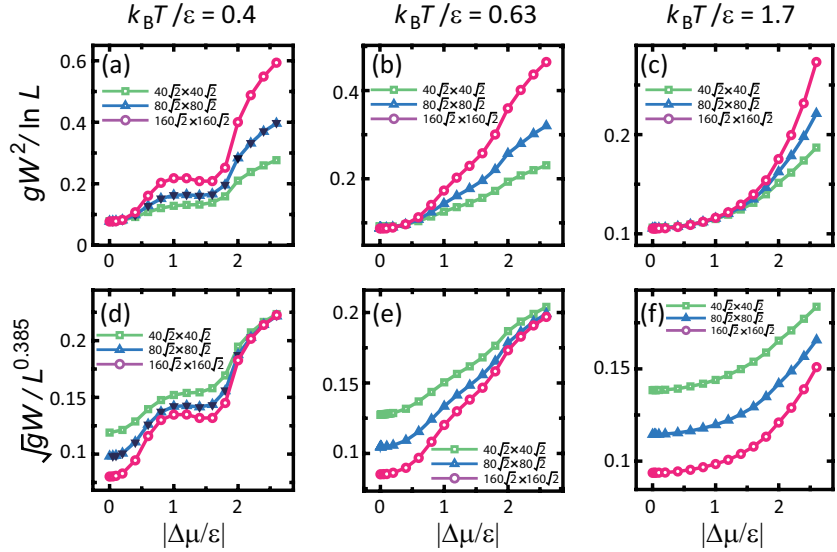}%
\caption{
Driving force dependence of the surface width.
(a), (b), and (c) show $gW^2/\ln L$ {\it vs.} $\Delta \mu/ \epsilon$.
(d), (e), and (f) show $\sqrt{g}W/L^{0.385}$ {\it vs.} $\Delta \mu/ \epsilon$.
(a), (c), (d), and (f): surface slope $p=3\sqrt{2}/8 \approx 0.530$, tilt angle $\theta$ = 27.9$^\circ$. 
(b) and (e): surface slope $p=\sqrt{2}/2 \approx 0.707$, tilt angle $\theta$ = 35.3$^\circ$.
Reverse triangles in (a) and (d): $\Delta \mu$ negative and  $L=80\sqrt{2}a$ ($a=1$).
}  
\label{w2-dmut04}
\end{figure}


\begin{figure}
\centering
\includegraphics[width=11 cm,clip]{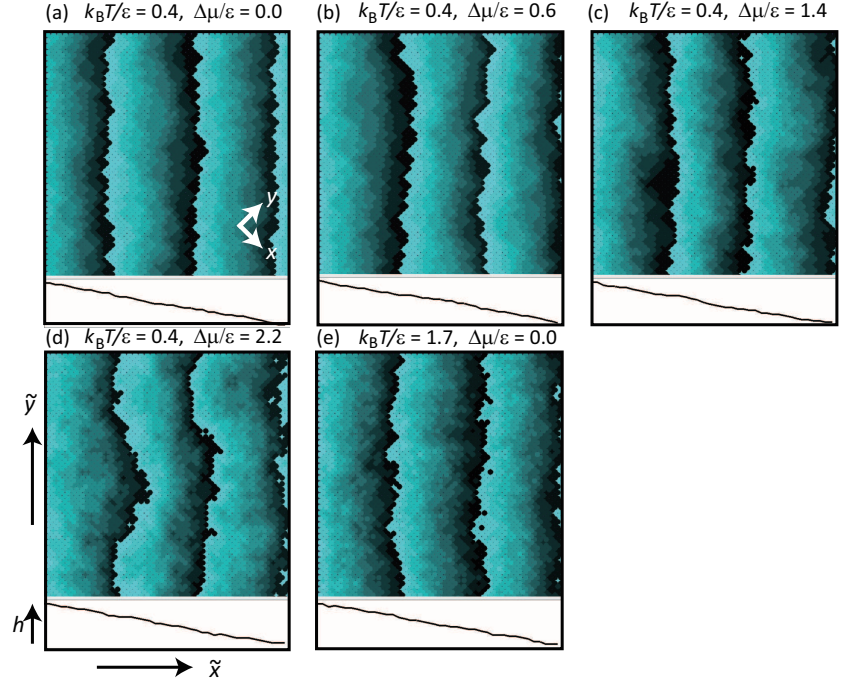}%
\caption{
Snapshots of simulated surfaces at $4 \times 10^8$ MCS/site. 
(a) and (e): BKT-rough surfaces.
(b), (c), and (d): KPZ-rough surfaces. 
Size: $40\sqrt{2}\times 40\sqrt{2}$.
$N_{\rm step}$ = 30.
$p =N_{\rm step}a/L =3\sqrt{2}/8\approx 0.530$. $\theta =27.9$ degree.
The surface height is represented by brightness with 10 gradations, where brighter regions are higher.
Due to the finite gradation, where the darkest areas sit next to the brightest areas, the darker area is higher by one gradation unit. 
The lines showing the side views are drawn with respect to the height along the bottom edge of the top-down views.
}  
\label{surfdat30}
\end{figure}

\subsection*{$\Delta \mu$ and $T$ dependence}

Figure \ref{w2-dmut04} shows the $|\Delta \mu|$ dependence of the surface width for several temperatures.
The roughening temperature of the (001) surface is $\TR^{(001)}/\epsilon = 1.55 \pm 0.02$ \cite{akutsu11JPCM,dennijs}, whereas the roughening temperature of the (111) surface $\TR^{(111)}$ is infinite.
The temperature in (c) and (f) is higher than $\TR^{(001)}$.

Near equilibrium ($\Delta \mu \sim 0$), the values of $gW^2/\ln L$ for each system size coincide  (Fig. \ref{w2-dmut04} (a), (b), and (c)), whereas for large $|\Delta|/\mu$, $\sqrt{g}W/L^\alpha$ with $\alpha=0.385$ for each system size converge as the driving force increases (Fig. \ref{w2-dmut04} (d), (e), and (f)).  
The value of $\alpha=0.385$ is the KPZ-exponent in this article.
Then, a crossover-driving-force $\Delta \mu_{co}$ is introduced. 
For $|\Delta \mu| < \Delta \mu_{co}$, $W^2 \propto \ln L$ (BKT rough), whereas for $\Delta \mu_{co}< |\Delta \mu| $, $W \propto L^\alpha$ (algebraic rough). 
From Fig. \ref{w2-dmut04}, it is clear that the value of $\Delta \mu_{co}$ depends on temperature. 
For high $|\Delta mu|$, the convergence of $\alpha$ to the KPZ value is stronger when the temperature is lower. 
At $\kBT/\epsilon=0.4$, $\Delta \mu_{co}|_{k_{\rm B}T/\epsilon=0.4} = 0.3 \epsilon$ (Fig. \ref{w2-dmut04} (a)).
For $|\Delta \mu/\epsilon|>2$, there is good agreement between the three lines for $\sqrt{g}W/L^{0.385}$ (Fig. \ref{w2-dmut04} (d)).

It has been suggested that a KPZ-rough surface may appear when the surface is kinetically roughened \cite{wolf91,barabasi,pimpinelli} because islands on the terraces enhance the step-growth velocity.
Hence, the driving force for the kinetic roughening $\Delta \mu_{kr}^{(001)}$ on the (001) surface is studied.
The obtained $\Delta \mu_{kr}^{(001)}$ at $\kBT/\epsilon=0.4$ is 
$\Delta \mu_{kr}^{(001)}/\epsilon=1.15 \pm 0.15$.
$\Delta \mu_{kr}^{(001)}$ is determined as follows.
For a smooth terrace surface, the surface velocity $V$ on the (001) surface converges to zero as the surface slope $p \rightarrow 0$; whereas, for a rough terrace surface, $V$ on the (001) surface converges to a finite value as  $p \rightarrow 0$.
Then, $\Delta \mu_{kr}^{(001)}$ is determined as the largest $|\Delta \mu|$, so that the surface velocity $V$ converges to zero as the slope $p \rightarrow 0$ \cite{wolf91} (refer to the section on the surface velocity below).
In Fig. \ref{w2-dmut04} (a) and (d), $W$ near $\Delta \mu_{kr}^{(001)}$ (around $\Delta \mu/\epsilon =1$) seems to form a broad peak for larger system sizes.
This peak in $W$ is considered to relate to the kinetic roughening of the (001) surface.
In snapshots of the surface (Fig. \ref{surfdat30}), the steps rarely have an overhang structure (Fig. \ref{surfdat30} (a) and (b)) for $\Delta \mu<\Delta \mu_{kr}$, whereas overhang structures on the contour lines can be seen for Fig. \ref{surfdat30} (c) and (d).
Figure \ref{surfdat30} (e) shows the thermally roughened surface.

It is interesting that the kinetic roughening occurs approximately  where the linear size of the 2D critical nucleus on a (001) terrace is less than $2a$ ($a=1$), where $a$ is the lattice constant.
Assuming that  the shape of the critical nucleus on a (001) terrace is square, the size of the  critical nucleus $r^*$ is expressed by $r^*/a=2\epsilon/\Delta \mu$.
At $|\Delta \mu/\epsilon|=1$ or 2, $r^*/a=2$ or 1, respectively.
Islands with a compact shape are frequently formed near $|\Delta \mu/\epsilon| = 1.0$ and merge with the step on the same layer. 
The step edge consists of several 1D ``overhang'' structures due to merging of the islands with steps.
In this manner, islands on a terrace enhance the step velocity.
For $r^*/a \leq 1$, where $\Delta \mu/\epsilon \geq 2$, even a single atom on the terrace grows to form an island.
This relates to the fact that $W$ increases drastically around $\Delta \mu/\epsilon = 2$.

It should be noted that the TSK picture is broken for $|\Delta \mu| \geq \Delta \mu_{kr} $ or for $T>\TR$, since the ``step'' is not well-defined due to  the terrace being roughened.
However, the contour lines on the surface shown in Fig. \ref{surfdat30} (c), (d), and  (e) (and Supplementary Fig. S1 (b), (c), and (d)) show the complexities of the surface.  
In  Fig. \ref{surfdat30} (d) and Supplementary Fig. S1 (c),  dendritic contour shapes can be seen.

In the case of crystal recession, the 2D nucleus on the (001) terrace at $\Delta \mu/\epsilon= -1$ is a {\it negative} square nucleus.
Here, an ad-hole, a negative-island, and a negative-nucleus are, respectively, a vacancy on the terrace, an island made by a vacancy, and a negative-island with a critical size.

At $\kBT/\epsilon = 0.63$, the characteristics of the $|\Delta \mu|$ dependence of $W$ are similar to those at $\kBT/\epsilon = 0.4$.
We also have $\Delta \mu_{co}|_{k_{\rm B}T/\epsilon=0.63} = 0.5 \epsilon$, whereas $\Delta \mu_{kr}|_{k_{\rm B}T/\epsilon=0.63} = 0.65\epsilon \pm 0.05\epsilon$.
The value of $W$ is smaller than that for $\kBT/\epsilon = 0.4$; the peak of $W$ around $\Delta \mu_{kr}|_{k_{\rm B}T/\epsilon=0.63}$ is small.

At $\kBT/\epsilon =1.7$ where $T>\TR^{(001)}$, the (001) terraces are rough at $\Delta \mu=0$.
Hence, there is no kinetic roughening.
Here, $\Delta \mu_{co}|_{k_{\rm B}T/\epsilon=1.7} = 1.2 \epsilon$, which is the largest among the three cases.
Multi-layer island formation caused by thermal fluctuations increases the region of BKT roughening (Fig. \ref{surfdat30} (e)).

\subsection*{$L$ Dependence}


\begin{figure}
\centering
\includegraphics[width=9cm,clip]{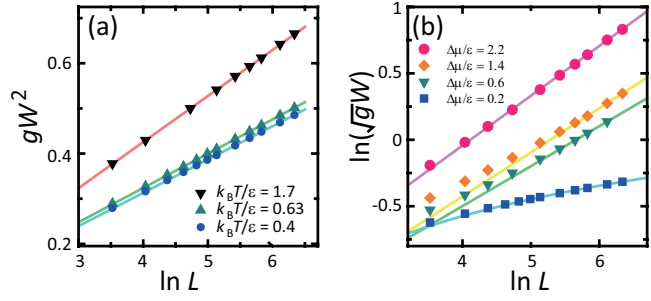}%
\caption{
System-size dependence of surface width.
$p=3\sqrt{2}/8 \approx 0.530$.
(a) $gW^2$ {\it vs.} $\ln L$ at $\Delta \mu = 0$.
Lines: from the top, $gW^2=0.0174+0.102 \ln L$, $gW^2=0.0201+0.0759 \ln L$, and $gW^2=0.0178+0.0738 \ln L$.
(b)  $\ln(\sqrt{g}W)$ {\it vs.} $\ln L$.  $\kBT/\epsilon =0.4$.
Lines: from the top, $\sqrt{g}W=0.214 L^{0.374}$, $\sqrt{g}W=0.175 L^{0.331}$, $\sqrt{g}W=0.182 L^{0.301}$, and  $gW^2=-0.0497+0.0917 \ln L$.
}  
\label{fig_gw2-lnL}
\end{figure}

Figure \ref{fig_gw2-lnL} shows the $\ln L$ dependence of $gW^2$ and $\ln (\sqrt{g}W)$.
Figure \ref{fig_gw2-lnL} (a) shows the results at equilibrium.
In contrast to the two-component system of our previous work at equilibrium \cite{akutsu20}, the linearity of the obtained data is high.
This indicates that the elementary steps are well separated and the intervals between kinks are small relative to the system size.
The amplitudes of the lines at $p= 3\sqrt{2}/8$ increase as the temperature increases, specifically, 0.102, 0.0759, and  0.0738 for $\kBT/\epsilon =1.7$, 0.63, and 0.4, respectively.
These amplitudes are larger than the universal value of $1/(2\pi^2) \approx 0.0507$ for $p \rightarrow 0$ \cite{akutsu88, akutsu88-1,yamamoto94}.

Figure \ref{fig_gw2-lnL} (b) shows results for the non-equilibrium steady-state at $\kBT/\epsilon=0.4$.
For small $\Delta \mu$, $gW^2$ increases logarithmically as the system size increases.
However, a power law behaviour of $\sqrt{g}W$ is obtained for relatively large $|\Delta \mu|$.
The slopes of the lines show the roughness exponent $\alpha$.
For large $L$, the obtained $\alpha$ for $\Delta \mu/\epsilon=2.2$, $1.4$, and $0.6$ are $0.347$, $0.331$, and $0.301$, respectively.
This is consistent with the results seen in Fig. \ref{w2-dmut04} (d)--(f).
The exponent $\alpha$ seems to gradually increase as $|\Delta \mu|$ increases.
However, the slope at larger $L$ is steeper.
Therefore, we consider that the exponent $\alpha$ converges to the KPZ value in the limit of $L \rightarrow \infty$.
The large finite size effect decreases the value of $\alpha$ in the small length region.

It is interesting that large wavelength surface fluctuations are observed in the snapshots in Fig. \ref{surfdat30} (b), (c), and (d).
We also show snapshots for $L=400\sqrt{2}a$ in the Supplementary Information.

From the results in this and the previous sections, we conclude that the crossover point $\Delta \mu_{co}$ between the BKT-rough and the algebraic-rough surfaces is different from the kinetic roughening point $\Delta \mu_{kr}$.
Also, the algebraic-rough surface is essentially the KPZ-rough surface in the limit of $L \rightarrow \infty$.
The large finite size effect decreases the value of $\alpha$ for the small system size.

\subsection*{$p$ dependence}


\begin{figure}
\centering
\includegraphics[width=9 cm,clip]{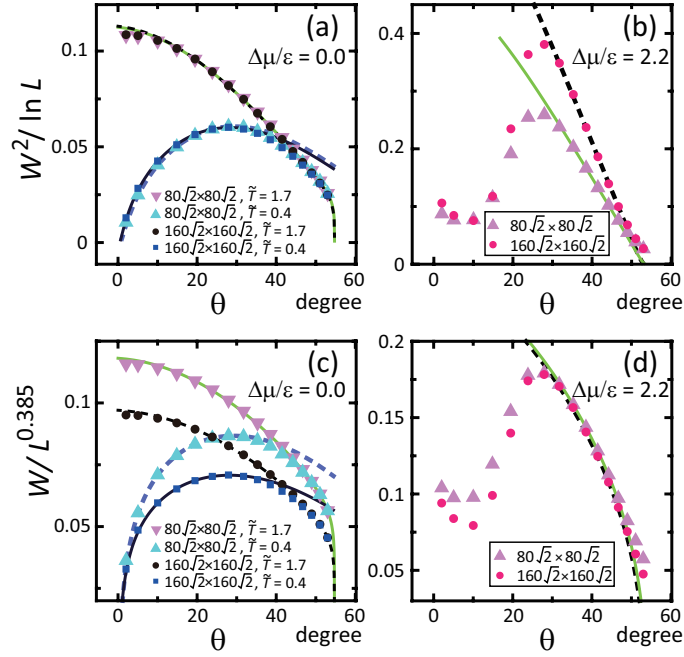}%
\caption{
Slope dependence of $W$.
$p=\tan \theta$.
$g= 1+p^2$.
$\tilde{T}=\kBT/\epsilon$.
(a) and (c) Dark solid line: $\sqrt{g}W=(0.317+ 0.0622 \ln p)\sqrt{\ln L}$.
Light broken line: $\sqrt{g}W=(0.321+ 0.0677 \ln p)\sqrt{\ln L}$.
Light solid line: $\sqrt{g}W=[0.327+ 0.0249 \ln (\sqrt{2}-p)]\sqrt{\ln L}$.
Dark broken line: $\sqrt{g}W=[0.327+ 0.0269 \ln (\sqrt{2}-p)]\sqrt{\ln L}$.
(b) and (d) $\kBT/\epsilon = 0.4$.
Light solid line: $\sqrt{g}W=[0.233+ 0.0836 \ln (\sqrt{2}-p)]L^{0.374}$.
Dark broken line: $\sqrt{g}W=[0.233+ 0.0886 \ln (\sqrt{2}-p))]L^{0.374}$.
}  
\label{scaled}
\end{figure}

Figure \ref{scaled} shows the slope dependence of $W$ and $W^2$.
At equilibrium, apart from the neighbourhood of the (001) surface, $W$ for $\kBT/\epsilon=0.4$ is well described by (Fig. \ref{scaled} (a)) the following single equation: 
\beq
gW^2/(\ln L) = (A + B \ln p)^2, \quad 
A=0.319\pm 0.006, \ B= 0.065 \pm 0.008. \label{pdepend_lowp}
\eeq
Here, for $T> \TR^{(001)}$,  $gW^2/\ln L$ of the (001) surface converges to a finite value for $p \rightarrow 0$.
For different temperatures for $T<\TR^{(001)}$, the slope dependence of $W$ agrees well within 5\%. 

For a large surface slope, near the (111) surface where $\TR^{(111)}$ is infinite, $W$ is well described by the  following single equation:
\beq
gW^2/(\ln L) = [A' + B' \ln (\sqrt{2}-p)]^2,  \quad 
A'= 0.327 \pm 0.002, \ B' = 0.026 \pm 0.005, \label{pdepend_highp}
\eeq
for $\kBT/\epsilon =$ 0.4, 0.63, and 1.7.
It should be noted that the vicinal surface around the (111) surface of the RSOS model is approximate compared to the real (111) surface.
The negative-step (Fig. \ref{schem}) is a ``step'' with a (111) terrace  in the step-down direction.
 However, due to the geometrical restrictions of the model, there are no ``steps'' with a (111) terrace  in the step-up direction.
For the same reason, there are no ad-atoms or ad-holes on the (111) terraces either.

For the non-equilibrium steady state, from Fig. \ref{scaled} (d), apart from the neighbourhood of the (111) surface, the slope dependence of $W$ is well described by
\beq 
gW/L^{\alpha}=A'' + B'' \ln (\sqrt{2}-p), \quad 
A''= 0.233 \pm 0.002, \ B''= 0.086 \pm 0.004, 
\quad \alpha=0.374,
\eeq
at $|\Delta \mu/\epsilon|= 2.2$ and $\kBT/\epsilon =0.4$.

Unexpectedly, as seen from Fig. \ref{scaled} (b) and (d) for small $\theta$, a vicinal surface with a small tilt angle shows a different behaviour from the KPZ-rough surface even if $|\Delta \mu|$ is high.
For $\theta< 19^\circ$, the vicinal surface is BKT-rough.
We will return to this point in the discussion.

\subsection*{Mean height of locally merged step}


\begin{figure}
\centering
\includegraphics[width=10.0 cm,clip]{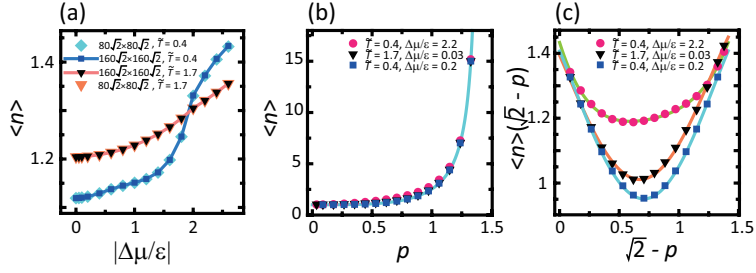}%
\caption{
$|\Delta \mu|$ and the slope dependence of the mean height of a locally merged step $\langle n \rangle$.
$\tilde{T}=\kBT/\epsilon$.
(a) $|\Delta \mu|$ dependence of $\langle n \rangle$.
$p=3\sqrt{2}/8 \approx 0.530$.  $\theta = 27.9^\circ$.
(b) Slope dependence of $\langle n \rangle$.
(c)  Slope dependence of $\langle n \rangle(\sqrt{2}-p)$.
 Lines: $\langle n \rangle (\sqrt{2}-p)  = A+B(\sqrt{2}-p-C)^2+D(\sqrt{2}-p-C)^3+E(\sqrt{2}-p-C)^4$; from top to bottom, $A=1.19$, $B= 0.290$, $C=0.594$, $D=-0.352$, and $E=0.535$; $A=1.01$, $B= 1.07$, $C=0.672$, $D=0$, and $E=-0.493$; and $A=0.951$, $B= 1.21$, $C=0.707$, $D=0$, and $E=-0.604$. 
}  
\label{nav}
\end{figure}

Figure \ref{nav} (a) shows the $\Delta \mu$ dependence of the mean height of locally merged steps $\langle n \rangle$.
In contrast to the cases of surfaces with faceted macrosteps \cite{akutsu18, akutsu19, akutsu19-2}, $\langle n \rangle$ is independent of the system size or the initial configurations.
This lack of a finite size effect  means that $\langle n \rangle$ in the RSOS model is determined by the local  or short wavelength  structure of steps.

It is interesting that $\langle n \rangle$ at $\kBT/\epsilon =0.4$ increases rapidly  for $\Delta \mu/\epsilon>1.2$, which is almost the same as $\Delta \mu_{kr}$ at $\kBT/\epsilon=0.4$.
At equilibrium, the result that $\langle n \rangle \sim 1$ indicates that the steps are well separated.
When $|\Delta \mu|$ is about $\Delta \mu_{kr}$, 2D nucleation with a compact shape and growth occurs frequently on the (001) terraces (Fig. \ref{surfdat30} (c), Supplementary Fig. S1 (b)).
The growing islands merge with the step on the same layer to enhance the surface growth velocity.
However, growing islands that catch up with steps on the lower layer are prevented from further growth due to geometrical restrictions.
Hence, the ratio of multi-height steps increases.
This is why $\langle n \rangle$ increases rapidly as $|\Delta \mu|$ increases for $|\Delta \mu|> \Delta \mu_{kr}$.
Assuming that the increase of $\langle n \rangle$ is dominantly caused by the formation of double steps, the ratio of the double step is less than 20\% for $\Delta \mu/\epsilon \leq 1.6$, whereas the ratio increases up to about 50\% for $\Delta \mu/\epsilon > 1.8$ as $\Delta \mu$ increases.

Again, it should be noted that when $|\Delta \mu|$ exceeds $\Delta \mu_{kr}$, the TSK picture breaks down.
However, regarding a contour line on the surface as an extended meaning of a ``step'', the complexity of the surface undulations can be explained by an extended T``S''K picture.

At high $|\Delta \mu|>2$, since the size of the critical nucleus  is less than one, ad-atoms on the terrace frequently grow larger islands  for crystal growth.
Also, the ad-atoms rarely escape from the terrace and the islands have dendrite shapes.
Hence, by merging to a step, the contour lines of the vicinal surface exhibit winding shapes (Fig. \ref{surfdat30} (d), Supplementary Fig. S1 (c)).

The slope dependence of $\langle n \rangle$ was also calculated (Fig. \ref{nav} (b)).
$\langle n \rangle$ is approximated by $\langle n \rangle \approx A\sqrt{2}/(\sqrt{2}-p)$ with $A=0.672$ for $\kBT/\epsilon =0.4$ and $\Delta \mu/\epsilon= 0.2$.
More precisely, $\langle n \rangle$ is relatively large near the (001) and (111) surfaces (Fig. \ref{nav} (c)).
For a BKT-rough surface, $\langle n \rangle(\sqrt{2}-p)$ is well expressed by a quadratic function with respect to $(\sqrt{2}-p)$; whereas for a KPZ-rough surface, $\langle n \rangle(\sqrt{2}-p)$ is asymmetric around $p=1/\sqrt{2}$.

\subsection*{Surface velocity\label{sec_surf_velocity}}


\begin{figure}
\centering
\includegraphics[width=12.0 cm,clip]{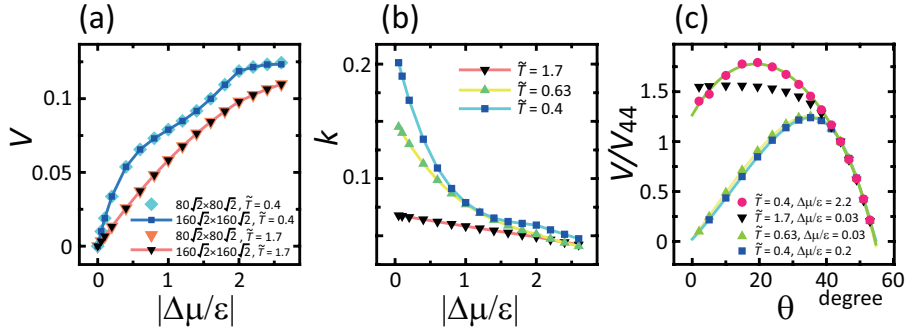}%
\caption{
(a) Surface growth velocity $V$.
 $p=3\sqrt{2}/8 \approx 0.530$.  $\theta = 27.9^\circ$.
$\tilde{T}=\kBT/\epsilon$.
(b) Kinetic coefficient $k = V \epsilon \tau/(a\Delta \mu)$. 
 $p=3\sqrt{2}/8 \approx 0.530$ for $\tilde{T}=1.7$ and $\tilde{T}=0.4$.
$p=\sqrt{2}/2 \approx 0.707$ for $\tilde{T}=0.63$.  $\theta = 35.3^\circ$.
(c) Slope dependence of the relative surface velocity.
$V_{44}$ is the surface velocity at $\theta =44^{\circ}$.
Lines:  $V/V_{44} = A+B(p-C)^2 +D(p-C)^3 +E(p-C)^4$, $p=\tan \theta$; from top to bottom, $A=1.78$, $B= -3.26$, $C=0.345$, $D=2.68$, and $E= -1.03$; $A=1.25$, $B= -3.26$, $C=0.694$, $D=0$, and $E= 1.46$; $A=1.24$, $B= -3.45$, $C=0.704$, $D=0$, and $E= 2.01$.
}  
\label{gr}
\end{figure}

Figure \ref{gr} (a) shows the $\Delta \mu$ dependences of the surface velocity $V$.
The surface velocity does not depend on the system size,
but is determined by the local structure of the surface, such as the kink density on the surface.
To determine the $|\Delta \mu|$ dependence of the kink density, the kinetic coefficient $k= (V/\Delta \mu)(\epsilon\tau/a)$ \cite{chernov61} was calculated, where $\tau$ is the interval time of one MCS/site (Fig. \ref{gr} (b)).
Unexpectedly, for $T<\TR$, the kinetic coefficient $k$ decreases rapidly as $|\Delta \mu|$ increases up to $\Delta \mu_{kr}$; $k$ decreases gradually for $|\Delta \mu|>\Delta \mu_{kr}$.
The change of $k$ happens in the step flow region rather than in the kinetically roughened region.
For $T> \TR^{(001)}$, $k$ decreases by a constant rate as $|\Delta \mu|$ increases.

Considering the surface velocity, a stepwise increase can be seen in Fig. \ref{gr} (a) for $\kBT/\epsilon= 0.4$.
The step flow growth is almost saturated around $\kBT/\epsilon \sim 1$.
For $|\Delta \mu/\epsilon| > \Delta \mu_{kr}$, regarding the contour lines as extended surface ``steps'', additional surface growth (or recession for $\Delta \mu <0$) occurs by a 2D dendritic-island-growth process.
Hence, the surface growth shows a stepwise increase.
A stepwise increase of $V$ with respect to $\Delta \mu$ is also observed experimentally for a metal-alloy surface \cite{Zhang}. 
It is known that the islands on the terrace surface have dendritic shapes, which is consistent with the present observation of the contour shapes in the computer simulations (Fig. \ref{surfdat30} (d), Supplementary Fig. S2 (c)).

Figure \ref{gr} (c) shows the slope dependence of the relative surface velocity.
$V_{44}$ is the surface velocity of the surface with $\theta = 44^{\circ}$.
$V/V_{44}$ is equal to $k/k_{44}$, where $k_{44}$ is the kinetic coefficient of the surface with $\theta = 44^{\circ}$.
In the limit of $\theta=0$, $V/V_{44}$ for a thermally rough (001) surface and a kinetically rough (001) surface converge to finite values; whereas  for a smooth (001) surface,  $V/V_{44}$ converges to zero.
Using these characteristics, $\Delta \mu_{kr}$ can be determined.

It is remarkable that, for $\theta > 42^{\circ}$ or $p> 0.90$, the slope dependence of $V/V_{44}$ coincides well with one of the curves, regardless of the difference of $T$ or $\Delta \mu$. 
This indicates that the local structure or the short range structure of the vicinal surface for $\theta > 42^{\circ}$ is approximately the same.

\section*{Discussion}

The relationship between the surface velocity and the fluctuation width was discussed by Wolf \cite{wolf91} using the renormalization group method.
Let us consider the anisotropic KPZ  (AKPZ) model \cite{wolf91, villain91}:
\beqa
\frac{\partial h}{\partial t} &=&  \nu_{\tilde{x}} \frac{\partial^2 h}{\partial \tilde{x}^2} 
+ \nu_{\tilde{y}} \frac{\partial^2 h}{\partial \tilde{y}^2} 
+ \frac{\lambda_{\tilde{x}}}{2}(\frac{\partial h}{\partial \tilde{x}})^2
+\frac{\lambda_{\tilde{y}}}{2}(\frac{\partial h}{\partial \tilde{x}})^2+ \eta(\vec{x},t), \nonumber \\ 
&& \langle \eta(\vec{x},t) \rangle = 0,   
\quad \langle \eta(\vec{x},t)\eta(\vec{x}',t') \rangle 
  = D\delta(\vec{x}-\vec{x}')\delta(t-t'), 
\label{eq_kpz} 
\eeqa 
where $\nu_{\tilde{x}}$ ($\nu_{\tilde{x}}$) is a relaxation constant for the $\tilde{x}$ ($\tilde{y}$) direction related to the surface tension,  $\lambda_{\tilde{x}}$ ($\lambda_{\tilde{y}}$) is the coefficient related to the ``excess velocity'', and $\eta(\vec{x},t)$ is Gaussian white noise.
The parameters $\lambda_{\tilde{x}}$ and $\lambda_{\tilde{y}}$ are given by
\beq
\lambda_{\tilde{x}}=\partial^2 V/\partial p^2, 
\ \lambda_{\tilde{y}}=(\partial V/\partial p)/p. \label{eq_lambda}
\eeq
Wolf \cite{wolf91} found that for $\lambda_{\tilde{x}} \lambda_{\tilde{y}} > 0$, the system converges to a fixed point with  algebraic roughness ($W \propto L^\alpha$); whereas for $\lambda_{\tilde{x}}\lambda_{\tilde{y}}<0$, the system converges to another fixed point with logarithmic roughness  ($W^2 \propto \ln L$).

To compare our results to the AKPZ model, we investigate the consistency between our results and the AKPZ results.
For small $\theta$, $\partial V/\partial p >0$ and $\partial^2V/\partial p^2<0$.
Then, $\lambda_{\tilde{x}}\lambda_{\tilde{y}}<0$ indicates that the surface should be BKT (logarithmic)-rough.
This is consistent with our results.
For large $\theta$,  $\partial V/\partial p <0$ and $\partial^2V/\partial p^2<0$.
Then, $\lambda_{\tilde{x}}\lambda_{\tilde{y}}>0$ indicates that the surface is KPZ (algebraic)-rough.
This seems to be consistent with the results for the large $|\Delta \mu|$ case.
However, for small $|\Delta \mu|$, our results show BKT-roughness for large $\theta$. 
Hence, the AKPZ results are not fully consistent with our results.
More seriously, if the surface slope $p$ is replaced by $\sqrt{2}-p$ and redefined by $\hat{p}$, then we have $\partial V/\partial \hat{p} >0$ and $\partial^2V/\partial \hat{p}^2<0$ for large $\theta$ surfaces, which leads to logarithmic roughness.
The AKPZ results change depending on the definition of the slope $p$.  
Therefore, the AKPZ results cannot be established for the case of large $\theta$ in our model.

Then, the question remains as to what is the ``relevant'' quantity to make a KPZ (algebraic)-rough surface from a BKT-rough surface.
We focused on the difference between the surfaces of $\kBT/\epsilon=0.4$ with $|\Delta \mu|=0$ and $|\Delta \mu/\epsilon|=2.2$ for $\theta > 30^\circ$.
The only difference between them in the external parameters is the value of $|\Delta \mu|$.
Therefore, we conclude that a sufficiently large $|\Delta \mu|>0$ creates a KPZ (algebraic)-rough surface.

The next question is why a large $|\Delta \mu|>0$ creates a KPZ (algebraic)-rough surface.
We consider that a sufficiently strong asymmetry between the attachment and detachment of atoms, which creates overhang structures on negative-step \cite{akutsu16} edges, gives rise to the KPZ-rough surface.
Here, a negative-step is a step such that the terrace is the (111) surface and the side surface of the negative-step is the (001) surface (Fig. \ref{schem}).
For large $\theta$, where the vicinal surface is close to the (111) surface, the surface grows or recedes at the edge of negative-steps on the (111) surface (Supplementary Fig. S2).
Due to the geometric restrictions of the RSOS model, ad-atoms or ad-holes are forbidden on the (111) surface. 
Hence, the (111) surface does not roughen  kinetically.
Nevertheless, the surface becomes a KPZ-rough surface when $|\Delta \mu|$ is sufficiently large.
Therefore, we conclude that a sufficiently strong asymmetry between attachment and detachment of atoms at negative-step edges creates the KPZ-rough surface.

Physically, for $|\Delta \mu/\epsilon|>>0$ with large $\theta$, growing negative-steps have a strongly anisotropic step velocity.
To describe this anisotropy, we introduce Miller indices for the (111) plane.
The $(01)$, $(10)$, and $(11)$ negative-steps have  2D vectors normal to the  mean step-running directions of  $\langle \bar{1}01 \rangle$, $\langle 0\bar{1}1 \rangle$, and $\langle 1\bar{1}0 \rangle$ directions, respectively.  
Since the step velocity of $(11)$ negative-steps is larger than that for $(01)$ or $(10)$ negative-steps, $(11)$ negative-steps with a small-scale zig-zag structure involving $(01)$ and the $(10)$ negative-steps under non-equilibrium conditions tend to be surrounded by longer $(01)$ and $(10)$ steps.
Then, larger square shapes with (01) and (10) negative-steps  than square shapes with (01) and (10) negative-steps at equilibrium are formed (Supplementary Fig. S2).
Some produce an overhanging structure at the negative-step edges. 
In this manner, a large scale zig-zag structure with overhangs on the negative-step edges is formed due to the anisotropy in the step velocity for $|\Delta \mu/\epsilon|>>0$, which increases the width of surface fluctuations.
Therefore, the anisotropy in the step velocity or of the kink density at the step edges  creates KPZ-roughness on the surface for $|\Delta \mu/\epsilon|>>0$.

The third question is why the vicinal surface with small $\theta$ shows a BKT-rough surface even for large $|\Delta \mu|$.
We consider that ad-atoms, ad-holes, islands, and negative-islands on ``terraces'' block the advancement/recession of the ``steps'', decreasing the surface fluctuation width. 
The only difference between a BKT-rough surface and a KPZ-rough surface is the surface slope for $\kBT/\epsilon=0.4$ and $\Delta \mu/\epsilon= 2.2$.
For a small $\theta$ surface, ad-atoms, ad-holes, islands, and negative-islands form on the (001) terrace (Fig. \ref{schem}, Fig. \ref{surfdat30} (b), (c), and (d)).
When these excitations exist on the same layer as a step, they help to grow/recede the step.
However, when such excitations exist on different layers from the step, they hinder  the advancement/recession of the step.
In this manner, the surface fluctuation is suppressed, decreasing $W$.
For a large $\theta$ surface,  ad-atoms, ad-holes, islands, and negative-islands cannot form on the (111) terrace due to geometrical restrictions.
The surface can grow/recede mainly by growing/receding steps and negative-steps.
A similar situation occurs in a 2D lattice gas on a surface.
The phase transition in the 2D lattice gas model belongs to the 2D Ising class.
However, due to the presence of islands with multiple heights on the surface, the roughening transition of the surface belongs to the BKT class.
Therefore, we conclude that the ad-atoms, ad-holes, islands, and negative-islands are relevant to making the BKT-rough surface.

\section*{Conclusions}

For the RSOS model with a discrete Hamiltonian under a non-equilibrium steady state without surface diffusion or volume diffusion: 
\begin{itemize} 
\item The crossover point $\Delta \mu_{co}$ between the BKT (logarithmic)-rough surface and the KPZ (algebraic)-rough surface is different from the kinetic roughening point $\Delta \mu_{kr}$.
\item A step flow growth or recession leads intrinsically to a KPZ-rough surface due to the anisotropic step velocity, where the anisotropy is caused by the crystal structure.
\item The ad-atoms, ad-holes, and their clusters on terraces, which block the step advancement and recession, are relevant for making the BKT-rough surface. 
\end{itemize}




\section*{Acknowledgements}
The author wishes to acknowledge Prof. K. Takeuchi and Prof. T. Einstein for their advice on the relationship between random matrices and crystal growth.
This work was supported by KAKENHI Grants-in-Aid (nos. JP25400413 and JP17K05503) from the Japan Society for the Promotion of Science (JSPS).
This work was supported in part by the Collaborative Research Program of Research Institute for Applied Mechanics, Kyushu University.

\section*{Author contributions statement}

N.A. conceived and conducted the calculations, and  analyzed the results.

\section*{Additional information} 
\textbf{Supplementary information} is available for this paper at a URL.\\
%
%
\noindent
\textbf{Competing financial interests:} The author declares no competing interests.

\end{document}